# Solving Wave Equations in the Space of Schwartz Distributions: The Beauty of Generalised functions in Physics.

**Luca Nanni**

Faculty of Natural Science, University of Ferrara, 44122 Ferrara, Italy

**Abstract**

*This paper concerns the study and resolution of wave equations in the space of Schwartz distributions. Wave phenomena are widespread in many branches of physics and chemistry, such as optics, gravitation, quantum mechanics, chemical waves and often arise from instantaneous sources represented by Schwartz distributions $f(x)$. Hence, there is a need to study the Cauchy problem in the space of generalised functions. Specifically, it has been proven that the instantaneous source $f(x)$ can always be represented as an appropriate sum of single point-like sources. Under this hypothesis, each wave equation with an instantaneous source $f(x)$ remains associated with an equation with a point-like source represented by a Dirac delta function $\delta(x)$. The solution to the associated equation is an elementary perturbation that propagates in space–time, defined as the fundamental solution. We proved that the solution to a wave equation with source $f(x)$ is given by the convolution product between one of the fundamental solutions and the generalised function $f(x)$ representing the instantaneous source. We investigated the physical and mathematical properties of three-dimensional, two-dimensional, and one-dimensional fundamental solutions. Notably, we proved that the three-dimensional solution described diffraction phenomena, whereas the other two described wave diffusion phenomena. Furthermore, we demonstrated that the transition from a diffractive to a diffusive regime occurs through the continuation of an ansatz generalised function. In this paper, we discuss possible applications to solid-state physics and the resolution of crystallographic structures.*

## 1. Introduction

Generalised functions are mathematical objects introduced to extend ordinary functions [1-3]. The need for such extensions arises in many mathematical and physics problems [4-5], such as the mass density of a point-like particle, the charge density of a double layer or a dipole, and the perturbation propagating in deforming interfaces [6-7], which can only be addressed by generalising the classical concept of a function. From a physical perspective, the meaning behind a generalised function is that a certain quantity cannot be measured at a point; only its average value in a small but finite neighbourhood of that point can be determined.

Historically, the first hint of generalisation of an ordinary function came from Euler [8]. However, Euler's idea was ahead of its time, and mathematics still had no adequate tools for approaching such a problem. No progress was made until the nineteenth century when Heaviside introduced what is considered the first generalised function of mathematical physics: the step function [9]. A rigorous approach to distribution theory emerged only with the advent of quantum mechanics, inspired by the introduction of the Dirac delta function [8]. Within a few years, authoritative mathematicians developed what is now known as distribution theory [11–17], which is now one of the most widespread mathematical analyses applied to science. Specifically, in the second half of the twentieth century, it contributed significantly to formulating quantum field theory [18–21].

In this article, a wave equation describing phenomena caused by a generic instantaneous source is solved in the space of Schwartz distributions. Dirac once said, 'God used beautiful mathematics in creating the world'. Based on this famous quotation, we believe that the topic of this paper represents a case of the beauty of a physical theory meeting the elegance of mathematics. Wave theory is of interest in many branches of classical physics, chemistry, optics, and quantum mechanics. In many cases, the source acts instantaneously, nullifying the resolution of equations in the space of ordinary functions. However, their solutions can be recovered by extending ordinary functions in the space of Schwartz distributions. This paper shows that the initial source, represented by a generic generalised function $f(x)$, can be seen as the sum of the contributions of single point-like sources. The initial equation is thus associated with an equation for which the source is a Dirac delta function, and this equation can be solved by applying the Fourier transform ansatz. We define these solutions herein as fundamental solutions. We also demonstrate that once the fundamental solution for $\mathbb{R}^{n+1}$ is known, the one for $\mathbb{R}^n$ can be obtained as a continuation of the first in a low-order subspace. The solution to a wave equation with source $f(x)$ is obtained by the convolution product of the latter times one of the fundamental solutions of the associated equation, providing one-, two-, and three-dimensional fundamental solutions. We investigated the mathematical properties of these functions to determine their physical meanings and their application to solid-state physics and crystallography. Notably, in this paper, we make an analogy between the Patterson method used to resolve crystalline molecular structures containing heavy atoms and the formalism proposed in this study.

This approach to the fundamental solution of the multidimensional wave equation is simpler and more immediate than those already present in the literature [11,13]. The problem's physical-mathematical structure allows the use of the Fourier transform method, and the finite source can be constructed as a convolution of point sources mathematically represented by Dirac distributions. In this way, it is possible to bring out the physical meaning of the solution straightforwardly.

## 2. Preliminary Notions of Schwartz Distributions

In this section, we list some definitions of Schwartz distributions and the associated operations to facilitate understanding of the subsequent sections. For a rigorous examination of the subject, please see [1–5].

**Definition 1**. The defined space of test functions, denoted by $\boldsymbol{\mathcal{D}}$, is the set of infinitely differentiable complex-valued functions $\varphi(x)$ for a non-empty subset $U \subset \mathbb{R}^n$ that has compact support. The support of function $\varphi(\mathrm{x})$, denoted by $supp(\varphi)$, is the smallest closed set containing all the points for which $\varphi(x) \neq 0$.

**Definition 2**. Each continuous linear function in the space $\boldsymbol{\mathcal{D}}(\mathbb{R}^n)$ of test functions is called a generalised function. The latter is said to be regular if it can be obtained from a locally integrable function $f(x)$ in $\mathbb{R}^n$ and the following integral holds:

$$(f, \varphi) = \int f(x)\varphi(x)dx \ \forall \varphi \in \boldsymbol{\mathcal{D}}(\mathbb{R}^n). \tag{1}$$

Any other generalised function that does not satisfy the integral of Equation 1 is called singular. The Schwartz distribution $f(x)$ can always be obtained by weak convergence of sequences with suitable ordinary parametric functions $\{f_\varepsilon(x)\}_{\varepsilon \in \mathbb{R}}$, known as mother functions. The function $f_\varepsilon(x)$ can be seen as a smooth approximation of $f(x)$. The set of all functions in the space $\boldsymbol{\mathcal{D}}(\mathbb{R}^n)$ defines the space of generalised functions, denoted by $\boldsymbol{\mathcal{D}}'(\mathbb{R}^n)$. Therefore, the space $\boldsymbol{\mathcal{D}}'(\mathbb{R}^n)$ is the dual space $\boldsymbol{\mathcal{D}}(\mathbb{R}^n)$.

**Definition 3**. The $n$th derivative of a generalised function $f$ is given by

$$(\partial^n f, \varphi) = (-1)^n (f, \partial^n \varphi). \tag{2}$$

The Heaviside step function $\Theta(x)$ and the Dirac delta function $\delta(x)$ are examples of infinitely differentiable generalised functions in the space $\boldsymbol{\mathcal{D}}(\mathbb{R}^n)$. Notably, $\Theta'(x) = \delta(x)$.

**Definition 4**. The primitive of a generalised function $f \in \mathcal{D}'$ is given by

$$\left(f^{(-1)}, \varphi\right) = (f, \varphi) + c \int \varphi(t)dt \ \forall \varphi \in \mathcal{D}, \tag{3}$$

where $c$ is a complex constant.

**Definition 5**. The Fourier transform of a generalised function $f \in \mathcal{D}'$ is the linear function that maps a test function $\varphi$ to the action of $f$ on the Fourier transform of $\varphi$. Using the mathematical formalism this definition reads

$$(\mathfrak{F}(f), \varphi) = (f, \mathfrak{F}[\varphi]), \forall \varphi \in \mathcal{D} \ and \ \forall f \in \mathcal{D}' \qquad (4)$$

A similar formula holds for the Fourier anti-transform. It is important to note that, due to how the test function is defined (see Definition 1), $\varphi$ always admits Fourier transform. All the properties of the Fourier transform that hold for ordinary functions also hold for generalised functions. A generalised function that admits Fourier transform is defined as a tempered distribution. The Heaviside function $\Theta(x)$ and the Dirac function $\delta(x)$ are tempered generalised functions, the Fourier transforms of which are $\mathfrak{F}[\Theta] = \left[\frac{1}{ik} + \pi\delta(k)\right]$ and $\mathbb{1}(k)$ respectively.

**Definition 6**. Let be two generalised functions $f, g \in \mathcal{D}'$ locally integrable on $\mathbb{R}^n$. Then, their convolution product reads

$$\omega = f * g = \int f(y)g(y-x)dy. \qquad (5)$$

It can be proved that the new generalized function $\omega$ is also locally integrable. The cases that ensure the local integrability of the generalized functions $g$ and $f$ are: 1) at least one is finite; 2) at least one is a compactly supported distribution; 3) both functions are integrable in $\mathbb{R}^n$. Concerning case 2), it must be noted that necessary and sufficient condition for a generalized function $f$ to have a compact support is the existence of its linear and continuous extension to $\mathcal{C}^\infty(\mathbb{R}^n)$. Finally, the relations $(f * g) = (g * f)$, $\mathfrak{F}(f * g) = \mathfrak{F}(f) * \mathfrak{F}(g)$, and $d^n(f * g)/dx^n = [(d^n f/dx^n) * g] = [f * (d^n g/dx^n)]$ hold.

## 3. The Cauchy Problem for a Wave Equation in the $\mathcal{D}'(\mathbb{R}^n)$ Space

Let the following be a linear differential equation:

$$\left[\frac{d^n}{dx^n} + c_1(x)\frac{d^{(n-1)}}{dx^{(n-1)}} + \cdots + c_n(x)\right]\omega(x) = f(x), \qquad (6)$$

where $x \in \mathbb{R}^n$, $c_i(x) \in \mathcal{C}^\infty(\mathbb{R}^n)$, and $f(x) \in \mathcal{D}'(\mathbb{R}^n)$. The function $\omega(x)$ is the perturbation generated by an instantaneous source $f(x)$. We clarify that $f(x)$ is a generic generalised function that does not necessarily represent a point-like source. We seek solutions $\omega(x)$ in the space $\mathcal{D}'^{(\mathbb{R}^n)}$. Based on this assumption, and using the formalism of the generalised function, Equation 6 can be rewritten as

$$\left(\hat{L}(x, \partial)\omega(x), \varphi(x)\right) = \left(f(x), \varphi(x)\right), \qquad (7)$$

where $\hat{L}(x, \partial)$ denotes the linear differential operator of order $n$, and $\varphi(x) \in \mathcal{D}(\mathbb{R}^n)$. Let us now suppose that the functions $c_i(x)$ are real constants, and that $f(x) = \delta(x)$. The equation

$$\left[\frac{d^n}{dx^n} + c_1 \frac{d^{(n-1)}}{dx^{(n-1)}} + \cdots + c_n\right] \eta(x) = \delta(x), \qquad (8)$$

is then called an associated equation with a point-like source, and its solution $\eta(x)$ is defined as a fundamental generalised function. Although $\delta(x) \neq 0$, the solution to Equation 8 is not unique. In fact, if $\eta_0(x)$ is a solution to the equation $\hat{L}(\partial)\eta(x) = 0$, where $\hat{L}(\partial)$ is the differential operator of Equation 8, then $\big(\eta_0(x) + \eta(x)\big)$ is also a solution to Equation 8. To prove this statement, let us apply the operator $\hat{L}(\partial)$ to the function $\big(\eta_0(x) + \eta(x)\big)$:

$$\begin{aligned} \hat{L}(\partial)\big(\eta_0(x) + \eta(x)\big) &= \hat{L}(\partial)\eta_0(x) + \hat{L}(\partial)\eta(x) = \hat{L}(\partial)\eta(x) \\ &= \delta(x). \end{aligned} \qquad (9)$$

If we limit the study to tempered generalised functions, then Equation 8 can be solved by the Fourier transform method. For this purpose, let us state the following proposition:

**Proposition 1**. For $\eta(x) \in \mathcal{D}'(\mathbb{R}^n)$ to be a solution to Equation 8, it is necessary and sufficient for equality $\mathcal{P}(-ik)\mathfrak{F}[\eta(x)] = \mathbb{1}(k)$ to be verified, where $k$ is the variable of the transformed function, $\mathcal{P}(-ik)$ is the polynomial $[c_0 + c_1(-ik) + \cdots + c_n(-ik)^n]$, $\mathfrak{F}[\eta(x)]$ is the Fourier transformed generalised function, and $\mathbb{1}(k)$ is the function that returns the number 1 for all $k$ values.

To prove this proposition, let us suppose that $\eta(x) \in \mathcal{D}'(\mathbb{R}^n)$ is a solution to Equation 8. By applying to both sides of Equation 8 the Fourier transform, one obtains:

$$\begin{aligned} \mathfrak{F}[\hat{L}(\partial)\eta(x)] &= \mathfrak{F}\left[\sum_{i=1}^{n} c_i \frac{d^i}{dx^i}\eta(x)\right] = \sum_{i=1}^{n} c_i(-ik)^i \mathfrak{F}[\eta(x)] \\ &= \mathfrak{F}[\delta(x)]. \end{aligned} \qquad (10)$$

Since $\mathfrak{F}[\eta(x)] = \mathbb{1}(k)$ [19–20], it also follows that $\sum_{i=1}^{n} c_i(-ik)^i \mathfrak{F}[\eta(x)] = \mathbb{1}(k)$, proving Proposition 1.

Proposition 1 reduces the problem of solving Equation 8 to calculating the roots of the polynomial equation $\mathcal{P}(-ik)\mathfrak{F}[\eta(x)] = \mathbb{1}(k)$. The latter can be rewritten as $\mathfrak{F}[\eta(x)] = \mathbb{1}(k)/\mathcal{P}(-ik)$, which is defined as the set $\mathcal{J}$ that does not contain zeros in polynomial equation $\mathcal{P}(-ik)$. Denoting $\mathcal{Z}$ by the set of zeros in $\mathcal{P}(-ik)$, if $\mathcal{Z} \neq \emptyset$, the solution for the polynomial equation will not be unique. This implies that even the Equation 8 solution will not be unique, confirming what we previously mentioned. Let us consider a simple but remarkable example, where $\mathcal{P}(-ik) = k$. In this case, the solutions for the polynomial are obtained from the following equation:

$$\left\{ \frac{1}{k + i\varepsilon}, \frac{1}{k - i\varepsilon}, p.v.\left(\frac{1}{k}\right) \right\}, \tag{11}$$

where $\varepsilon$ is a parameter, such that $\varepsilon \to 0^+$. The first two solutions based on Equation 11 are propagators encountered in quantum field theory [18], whereas the last solution is the Cauchy principal value, defined in [23]:

$$p.v.\left(\frac{1}{t}\right) = \lim_{\varepsilon \to 0} \int_{\mathbb{R}\setminus(-\varepsilon,\varepsilon)} \frac{1}{k} dk. \tag{12}$$

All the functions in Equation 11 are locally integrable Schwartz distributions that differ, at most, for a constant term $\delta(k)$, for which the Fourier anti-transform can be computed [18]. This example demonstrates that the fundamental solutions for Equation 8 are not unique but differ for a constant term given by the Fourier anti-transform of $\delta(k)$. The latter is given by $\mathfrak{F}^{-1}[\delta(k)] = sin(Ra)/\pi a$, where $R$ and $a$ are numerical constants that depend on the boundary conditions of the problem. Let us now state the following theorem [13, 24–25]:

**Theorem 1.** Let $f(x) \in \mathcal{D}'(\mathbb{R}^n)$ be a given function in Equation 6, $c_i(x)$ be constant functions, and $\eta(x) \in \mathcal{D}'(\mathbb{R}^n)$ be a fundamental solution to Equation 8. If the convolution product $\eta(x) * f(x)$ exists in $\mathcal{D}'(\mathbb{R}^n)$ space, then it yields the solution $\omega(x)$ for Equation 6.

This is the theorem underpinning our study and the entire article. Below, we provide proof of Theorem 1 to allow us to clarify some of the issues encountered when solving the wave equation for the cases $n = 1,2,3$. To this end, substituting the function $\eta(x) * f(x)$ in Equation 6, and recalling the derivative property of the convolution product given in Definition 6, we obtain

$$\hat{L}(\partial)\big(\eta(x) * f(x)\big) = \sum_{i=1}^{n} c_i \frac{d^i}{dx^i}\big(\eta(x) * f(x)\big)$$
$$= \left( \sum_{i=1}^{n} c_i \frac{d^i}{dx^i}\eta(x) \right) * f(x). \tag{13}$$

Since, for Equation 8, the term $\sum_{i=1}^{n} c_i \frac{d^i}{dx^i}\eta(x)$ is equal to $\delta(k)$, it follows that $\hat{L}(\partial)\big(\eta(x) * f(x)\big) = \delta(k) * f(x) = f(x)$, which proves the theorem.

This proof helps in understanding the physical meaning of the Cauchy problem. As clarified at the beginning of this paper, $f(x)$ is an instantaneous source but not necessarily point-like. However, it can be seen as the sum of the contributions of single point-like sources of type $f(\xi)\delta(x - \xi)$. An optical diffraction grating, a crystal lattice, and a layer of absorbed polarised molecules are examples of discrete structures represented by the sum of terms $f(\xi)\delta(x - \xi)$. It thus becomes clear that function $\omega(x) = \int f(\xi)\delta(x - \xi)d\xi$ is nothing but the superposition of elementary perturbations generated by point-like sources distributed in space according to a given rule. This is the *elegance* of Schwartz distribution formalism, which makes a physical theory *fascinating*.

## 4. Solving the Wave Equation in $\mathcal{D}'(\mathbb{R}^n)$ Space for $n = 1, 2, 3$

In this study, we were interested in solving Equation 6, where $\hat{L}(\partial)$ was the d'Alembert operator $\left(\nabla_n{}^2 - \frac{1}{u^2}\frac{\partial^2}{\partial t^2}\right)$ and $n = 1,2,3$. In this case, both the source and the solution for the equation are functions of $n$ spatial coordinates and of the time coordinate. We considered only the values of time $t \in \mathbb{R}^+$. The associated equation for which the fundamental solutions must be computed is as follows:

$$\left(\nabla_n{}^2 - \frac{1}{u^2}\frac{\partial^2}{\partial t^2}\right)\eta_n(t,x) = \delta_n(x)\delta(t). \qquad (14)$$

The generalised function $\eta_n(t,x) \neq 0$ only when $t > 0$. This suggests that the function $\eta_n(t,x)$ can be written as the direct product of the Heaviside function, depending on the time coordinate, and a generalised function, depending only on the spatial coordinates $x' = (ut - x)$; that is, $\eta_n(t,x) = \theta(t)\chi_n(x')$. Vladimirov proved that a solution for the differential equation $\left[\frac{d^2}{d\xi^2} + a^2\right]\chi(\xi) = \delta(\xi)$ is the function $\chi(\xi) = \theta(\xi)\frac{sin(a\xi)}{a}$, where, in our case, $a = u$ (the proof of this statement is provided by Vladimirov in section 6.6, chapter II of reference [13]). We believe this solution is a good starting point for solving Equation 14. Before going ahead, let us point out that the direct product of two generalized functions $f$ and $g$ acting on two different variables $x$ and $y$ is defined as $\left(f(x) \cdot g(y), \varphi(x,y)\right) = \left(f(x), \left(g(y), \varphi(x,y)\right)\right)$.

Let us consider the case $n = 3$ and apply the Fourier transform introduced in Section 2 to both sides of Equation 14 (see Definition 5):

$$\frac{\partial^2}{\partial t^2}\mathfrak{F}[\eta_3(t,x)] + u^2|k|^2\mathfrak{F}[\eta_3(t,x)] = \mathbb{1}(k)\delta(t), \qquad (15)$$

where $x \in \mathbb{R}^3$ and $t \in \mathbb{R}^+$. We then force $\mathfrak{F}[\eta_3(t,x)]$ to be equal to

$$\mathfrak{F}[\eta_3(t,x)] = \theta(t)\frac{sin(u|k|t)}{u|k|}. \qquad (16)$$

Substituting Equation 16 in Equation 15, one obtains

$$\frac{\theta''(t)sin(u|k|t)}{u|k|} + 2\theta'(t)cos(u|k|t) = \mathbb{1}(k)\delta(t), \tag{17}$$

where $\theta'(t) = \delta(t)$ and $\theta''(t) = \delta'(t)$. In $\mathbb{R}^3$, the Dirac function $\delta_3(R^2 - |x|^2)$, which can be seen as a thin layer on a sphere of radius $R = ut$, admits the remarkable Fourier transform [13]:

$$\mathfrak{F}[\delta_3(R^2 - |x|^2)] = 4\pi R \frac{sin(R|k|)}{|k|}. \tag{18}$$

Equation 18 allows us to easily solve Equation 17 by applying the Fourier anti-transform to all terms of its sides. Therefore, we obtain the following solution:

$$\eta_3(t,x) = \frac{\theta(t)}{4\pi ut}\delta(u^2t^2 - |x|^2). \tag{19}$$

The action of the function $\eta_3(t,x)$ on a test function $\varphi \in \mathcal{D}(\mathbb{R}^3 \times \mathbb{R}^+)$ is given by

$$(\eta_3(t,x), \varphi) = \frac{1}{4\pi u^2}\int\limits_0^\infty \frac{1}{t}\int\limits_\Sigma \varphi(t,x)d\Sigma dt, \tag{20}$$

where $\Sigma$ is a sphere of radius $R$. Equation 20 will be useful later in this section.

Let us now consider the case $n = 2$, where $x \in \mathbb{R}^2$. The solution is obtained by following the same approach used for the case $n = 3$. The only difference is that the function $sin(R|k|)/|k|$ in $\mathbb{R}^2$ is the following remarkable Fourier transform [13]:

$$\mathfrak{F}\left[\frac{\theta(R - |x|)}{(R^2 - |x|^2)^{1/2}}\right] = 2\pi \frac{sin(R|k|)}{|k|}, \tag{21}$$

where $R = ut$. Thus, we obtain the solution

$$\eta_2(t,x) = \frac{1}{2\pi u}\frac{\theta(ut - |x|)}{(u^2t^2 - |x|^2)^{1/2}}. \tag{22}$$

In $\mathbb{R}^2$, $(u^2t^2 - |x|^2)$ is the equation for a flat disc of radius $R$. Therefore, the generalised function $\theta(ut - |x|)/(u^2t^2 - |x|^2)^{1/2}$ can be seen as an elementary layer of such a disc.

To conclude, the case $n = 1$ is dealt with once again using the same method, considering, however, that in $\mathbb{R}$, the function $sin(R|k|)/|k|$ corresponds to the following remarkable Fourier transform [13]:

$$\mathfrak{F}[\theta(ut - |x|)] = 2\frac{sin(R|k|)}{|k|}, \tag{23}$$

Equation 23 allows us to obtain the following solution:

$$\eta_1(t,x) = \frac{1}{2u}\theta(ut - |x|). \tag{24}$$

We have therefore obtained the fundamental solutions that yield the functions $\omega(t,x)$, which are the solutions to the equation $\left(\nabla_n{}^2 - \frac{1}{u^2}\frac{\partial^2}{\partial t^2}\right)\omega_n(t,x) = f(t,x)$, where $n = 1,2,3$.

Let us prove that solutions $\eta_1(t,x)$ and $\eta_2(t,x)$ are obtained from function $\eta_3(t,x)$ by applying the continuation method. This technique is more laborious than the one discussed so far, but we believe it is worth investigating because it is useful when discussing the physical meaning of solution $\omega(t,x)$ given in Section 6. In this regard, we define the continuation of a generalised function as in [27].

**Definition 7**. A Schwartz distribution $\omega \in \mathcal{D}'(\mathbb{R}^{n+1})$ admits continuation for the test functions $\mathcal{D}(\mathbb{R}^n)$ if the limit $\lim_{k\to\infty}\big(\omega, \varphi(x)\varepsilon_k(\zeta)\big) = \big(\omega, \varphi(x)\mathbb{1}(\zeta)\big)$ holds. The function $\varepsilon_k(\zeta) \in \mathcal{D}(\mathbb{R})$ is an element of sequence $\{\varepsilon_k\}$ converging to unity as $k \to \infty$, whereas $\varphi(x)\mathcal{D}(\mathbb{R}^n)$. Therefore, the test function on which the distribution $\omega$ acts, whose support belongs to $\mathbb{R}^{n+1}$, is the ordinary product of a test function $\varphi$, whose support belongs to $\mathbb{R}^n$, and a parametric test function ($k$ denotes the numeric parameter) whose support belongs to $\mathbb{R}$. In other words, the variable $x$ belongs to $\mathbb{R}^n$, while $\zeta$ belongs to $\mathbb{R}$. The pair $(x, \zeta)$ is the variable with $n + 1$ components of $\varphi(x)\varepsilon_k(\zeta)$. The constraint to which the parametric function $\varepsilon_k(\zeta)$ is subject is expressed by the fact that the sequence $\{\varepsilon_k\}$ weakly converges to $\mathbb{1}(\zeta)$ as $k \to \infty$. This definition can be extended for continuations on a space of test functions with support belonging to $\mathbb{R}^{(n-m+1)}$. In this case, the test function will be the product of a function defined in a closed set of $\mathbb{R}^{(n-m)}$ times $(m + 1)$ parametric functions, each defined in $\mathbb{R}$, whose sequences weakly converge to $\mathbb{1}$ when the respective parameter tends to infinity.

This definition is reminiscent of the analytic continuation of a generalised function, but it has a distinctive meaning. The analytic continuation concerns a class of Schwartz distributions defined on a real axis that can be continued to holomorphic functions in the upper and lower complex half-planes [26–27]. Continuation, however, is the property of a generalised function $\omega$ defined in $\mathbb{R}^{n+1}$ acting in the space of the test functions defined in $\mathbb{R}^n$. We begin by demonstrating the continuation of the fundamental function $\eta_3(t,x_1,x_2,x_3)$ in space $\mathcal{D}(\mathbb{R}^2)$ that corresponds to the function $\eta_2(t,x_1,x_2)$. Using Equation 20 and setting $\varphi(x_1,x_2,x_3) = \lim_{k\to\infty}\psi(t,x_1,x_2)\varepsilon_k(x_3)$, we obtain

$$\left(\eta_3(t, x_1, x_2, x_3), \varphi(x_1, x_2, x_3)\right)$$
$$= \lim_{k \to \infty} \frac{1}{4\pi u^2} \int_0^\infty \frac{1}{t} \int_\Sigma \psi(t, x_1, x_2) \varepsilon_k(x_3) d\Sigma dt, \qquad (25)$$

where $\varphi \in \mathcal{D}(\mathbb{R}^3)$, $\psi \in \mathcal{D}(\mathbb{R}^2)$, and $\varepsilon_k(x_3) \to \mathbb{1}(x_3)$ lead to $k \to \infty$. The surface integral on the right-hand side of Equation 25 is calculated for sphere $\Sigma = \{x \in \mathbb{R}^3 : u^2 t^2 = |x_1^2 + x_2^2|^2 + x_3^2\}$. Since function $\psi$ does not depend on the $x_3$ coordinate, the surface integral can be replaced by the integral for circle $\partial = \{x \in \mathbb{R}^2 : ut > |x_1^2 + x_2^2|\}$:

$$\lim_{k \to \infty} \frac{1}{4\pi u^2} \int_0^\infty \frac{1}{t} \int_\Sigma \psi(t, x_1, x_2) \varepsilon_k(x_3) d\Sigma dt$$
$$= \frac{1}{2\pi u} \int_0^\infty \int_\partial \frac{\psi(t, x_1, x_2)}{(u^2 t^2 - |x_1^2 + x_2^2|^2)^{1/2}} \mathbb{1}(x_3) dx_1 dx_2 dt \qquad (26)$$
$$= \frac{1}{2\pi u} \int \frac{\Theta(ut - |x|)}{(u^2 t^2 - |x_1^2 + x_2^2|^2)^{1/2}} \psi(t, x_1, x_2) \mathbb{1}(x_3) dx_1 dx_2 dt.$$

Summarising, $\lim_{k \to \infty} \left(\eta_3, \psi(x)\varepsilon_k(x_3)\right) = \left(\eta_2, \psi(x)\mathbb{1}(x_3)\right)$ proves that $\eta_2(t, x_1, x_2)$ is the continuation of $\eta_3(t, x_1, x_2, x_3)$ in space $\mathcal{D}(\mathbb{R}^2)$. From the last integral of Equation 26, we can state that $\eta_2(t, x_1, x_2)\mathbb{1}(x_3) = [\eta_3(t, x_1, x_2, x_3) * \delta(x_1, x_2)]\delta(t)\mathbb{1}(x_3)$; that is, $\eta_2(t, x_1, x_2)$ is an elementary perturbation that does not depend on $x_3$, generated by the source $\delta(x_1, x_2)\delta(t)\mathbb{1}(x_3)$ concentrated along the $x_3$-axis. In this sense, $\eta_2 = \int_{-\infty}^\infty \eta_3 dx_3$. Therefore, this approach requires that the fundamental solutions be locally integrated.

Using this approach, the fundamental solution $\eta_1(t, x_1)$ is the continuation of solution $\eta_2(t, x_1, x_2)$ of the test function $\mu(x_1)\mathcal{D}(\mathbb{R})$:

$$\frac{1}{2\pi u} \int \frac{\Theta(ut - |x|)}{(u^2 t^2 - |x_1^2 + x_2^2|^2)^{1/2}} dx_2$$
$$= \frac{\Theta(ut - |x|)}{2\pi u} \int_0^{\left(u^2 t^2 - |x_1^2 + x_2^2|^2\right)^{1/2}} \frac{dx_2}{(u^2 t^2 - |x_1^2 + x_2^2|^2)^{1/2}} \qquad (27)$$
$$= \frac{\Theta(ut - |x|)}{2\pi u} \int_0^1 \frac{ds}{(1 - s^2)^{1/2}} = \frac{\Theta(ut - |x|)}{2u} = \eta_1(t, x_1).$$

## 5. Properties of the fundamental solutions $\boldsymbol{\eta_1}$, $\boldsymbol{\eta_2}$ and $\boldsymbol{\eta_3}$

The fundamental solutions for the associated wave equation are all locally integrable, formed by generalised functions $\Theta$ and $\delta$, which are also locally integrable [28]. The support of functions $\eta_1$ and $\eta_2$ are the closure of the cone bounded by the half-lines $ut = -x$ and $ut = x$. The support of function $\eta_3$ is the boundary of the cone. Geometrically, the form of $\eta_1$ at a given instant $t = \tau$ is a step that develops along the $x_1$-axis at a height $1/(2u)$ from it. The form of function $\eta_2$ is of a square root with a value at $|x| = 0$ that is $1/(2\pi u)$. It tends to infinity as $|x| \to ut$. Finally, the form of $\eta_2$ is a delta Dirac function at $|x| = ut$, with a height equal to $1/(4\pi u^2 t^2)$.

Generalised functions $\eta_n(t,x)$ with $n = 1,2,3$ belong to $\mathcal{C}(\mathbb{R}^+)$ with respect to variable $t$, and as $t \to 0^+$, the following limits hold:

$$a)\ \eta_n(t,x) \to 0\ ;\ b)\ \frac{\partial \eta_n(t,x)}{\partial t} \to \delta(t)\ ;\ c)\ \frac{\partial^2 \eta_n(t,x)}{\partial t^2} \to 0. \qquad (28)$$

Let us prove the property $a)$ using the explicit form of $\eta_3$. Using Equation 20, one obtains

$$(\eta_3(t,x),\varphi) = \frac{\Theta(t)}{4\pi u^2 t} \int\limits_\Sigma \varphi(t,x)d\Sigma = \frac{\Theta(t)t}{4\pi} \int\limits_{\Sigma'} \varphi(ut\zeta)d\zeta, \qquad (29)$$

where $d\Sigma = u^2 t d\zeta$. Since the function on the last side of Equation 29 is infinitely differentiable with respect to $t \in \mathbb{R}^+$, and tends to zero as $t \to 0^+$, it follows that $(\eta_3(t,x),\varphi) \to 0$, which implies that $\eta_3(t,x) \to 0$.

To prove the property $b)$, let us introduce the following differentiation operation [29]:

$$\left( \frac{\partial^k f_n(t,x)}{\partial t^k}, \varphi(x) \right) = \frac{d^k}{dt^k}\left( f_n(t,x),\varphi(x) \right) \forall k \in \mathbb{N}. \qquad (30)$$

Setting $k = 1$ and considering the case $f_n(t,x) = \eta_3(t,x)$, we obtain

$$\left( \frac{\partial \eta_3(t,x)}{\partial t}, \varphi(x) \right) = \frac{d}{dt}\left( \eta_3(t,x),\varphi(x) \right) = \frac{1+t}{4\pi} \int\limits_{\Sigma'} \varphi(ut\zeta)d\zeta. \qquad (31)$$

Since $t \to 0^+$, the right-hand side of Equation 31 tends towards $\varphi(0)$. Altogether, we therefore have $\left( \frac{\partial \eta_3(t,x)}{\partial t}, \varphi(x) \right) = \varphi(0)$, which is the action of $\delta(x)$ on the test function $\varphi(x)$. Generalising for $n = 1,2$, the limit $\frac{\partial \eta_n(t,x)}{\partial t} \to \delta(t)$ of property $b)$ is thus proven.

Finally, let us consider the property $c)$. Applying Equation 30 with $k = 2$ to $\eta_3(t,x)$, one obtains

$$\left(\frac{\partial^2 \eta_3(t,x)}{\partial t^2}, \varphi(x)\right) = \frac{d^2}{dt^2}\big(\eta_3(t,x),\varphi(x)\big)$$

$$= \frac{1}{2\pi}\frac{d}{dt}\int_{\Sigma'} \varphi(ut\zeta)d\zeta + \frac{t}{2\pi}\frac{d^2}{dt^2}\int_{\Sigma'} \varphi(ut\zeta)d\zeta. \tag{32}$$

Since $\varphi(x)$ is an even function infinitely differentiable with respect to $t \in \mathbb{R}^+$, and $\varphi'^{(ut\zeta)} = 0$ at $t = 0$, it follows that the last side of Equation 32 tends to zero as $t \to 0^+$. Property $c)$ is thus proved.

The properties listed in Equation 28 are proved for $n = 1,2$ following the same approach used for $n = 3$.

## 6. Physical interpretation of the fundamental solutions $\eta_1$, $\eta_2$ and $\eta_3$

Let us consider a source $f(t,x)$ of any form, such that $f(t,x) \neq 0 \ \forall t \in \mathbb{R}^+$, and $f(t,x) = 0 \ \forall t < 0$. We proved that the solution to Equation 6 with constant coefficients is given by $\omega_n(t,x) = \eta_n(t,x) * f(t,x)$. We also anticipated that $\omega_n(t,x)$ would be formed by the superposition of elementary perturbations generated by point-like sources, the interaction of which is governed by the structure of $f(t,x)$. These elementary sources are given by generalised functions of the type $f(\bar{t},\bar{x})\delta(x-\bar{x})\delta(t-\bar{t})$, where $(\bar{t},\bar{x})$ are points in the region $\Omega$ occupied by the source $f(t,x)$. Notably, $supp[f(t,x)] \subset \Omega$. By invoking the principle of the superposition of waves, we can affirm that the perturbation $\omega_n(t,x)$ generated by source $f(t,x)$ propagates in those points in the space given by $supp[\eta_n(t,x)] \cup \Omega$. Let us denote this space as follows:

$$\mathcal{S}(\Omega) = \cup_{(\bar{t},\bar{x})\in\Omega}\, supp[\eta_n(t-\bar{t},x-\bar{x})] \tag{33}$$

The structure of $supp[f(t,x)]$ determines whether the interaction between the elementary waves takes place and defines the modality of their propagation in space.

Let us study the case $n = 3$ in detail. The elementary perturbation $\eta_3(t,x)$ is generated by the point-like source $\delta(x)\delta(t)$. At instant $t > 0$, it will be concentrated on a spherical surface of radius $ut$ centered at $x = 0$. This surface expands over time with velocity $u$. By applying the principle of the superposition of waves, the perturbation $\omega_3(t,x)$ generated by source $f(t,x)$ propagates in the space formed by the points that constitute the boundary $u(t - \bar{t}) = (x - \bar{x})$ of the positive cones $\Gamma^+$. The vertices of these cones are located in the region $\Omega$ of the source. The space affected by perturbation $\omega_3(t,x)$ is therefore given as follows:

$$\mathcal{S}_3(\Omega) = \cup_{(\bar{t},\bar{x})\in\Omega}\, fr[\Gamma^+(\bar{t},\bar{x})]. \tag{34}$$

The value of the perturbation $\omega_3(t,x)$ at point $(t,x)$ is determined by the values of source $f(t,x)$ on the lateral surface of the cone, denoted by $\Gamma_0^+$ and already overcome by the wavefront. The value of $\omega_3(t, x)$ in $x$ at time $t > 0$ is defined by the values of $f(t, x)$ in the preceding instants given by

$$\bar{t} = t - \frac{|x - \bar{x}|}{u}. \tag{35}$$

The term $|x - \bar{x}|/u$ in Equation 35 is the retard time necessary for the perturbation to reach point $x$, starting from $\bar{x}$. We have just discussed the formulation of the Huygens–Fresnel principle [30]. In this framework, for example, the generalised function $f(t, x)$ can apply to a diffraction grating, where every single slit is represented by the point source $\delta(x)\delta(t)$. The geometry with which the single slits are positioned in the grating is determined by the structure of $f(t, x)$. If the diffraction grating were the reciprocal lattice of a crystal, then the single point-like sources would be the nodes (represented by the parallel planes of the direct grating) [31]. Simultaneously, the function $f(t, x)$ would represent their periodic spatial distribution. This is relevant because, once the mathematical form of the perturbation $\omega_3(t, x)$ is known (which can be deduced by diffractometric measurements), it is possible to obtain the function $f(t, x)$ by deconvolution of the product of $\eta_3(t, x) * f(t, x)$. From $f(t, x)$, one can thus obtain information on the periodic structure of the reciprocal lattice from which to calculate that of the direct lattice. This would be a new way of resolving the structure of crystals [32].

Let us suppose $f(t, x)$ given by

$$f(t, x) = \omega_0(x)\delta'(t) + \omega_1(x)\delta(t). \tag{36}$$

For $t > 0$, the perturbation is determined by the values of $\omega_0(x)$ and $\omega_1(x)$ on the surface of a sphere $S(x, ut)$, whereas for $t = 0$, the perturbation is concentrated in the region $\Omega$ containing $supp[\omega_0(x)]$ and $supp[\omega_1(x)]$. The perturbation will reach a point $x$ outside $\Omega$ at instant $t_0 = d/u$, where $d$ is the distance of $x$ from the source. Since source $f(t, x)$ is not point-like, we can define $d$ as the minimum distance of $x$ from the source, and $D$ as its maximum distance. The passage of the perturbation in $x$ will then have a duration given by $(D - d)/u$. Therefore, the advanced wavefront will pass at instant $t_0$, and the retarded one will pass at instant $t_1 = D/u$. Furthermore, the advanced wavefront will be the envelope of spherical surfaces $S = (x, ut)$ in which $x$ runs the points of region $\Omega$, and the retarded wavefront will be the inner envelope of these spherical surfaces.

Let us study the case $n = 2$. The elementary perturbation $\eta_2(t, x)$ is generated by a point-like source $\delta(x)\delta(t)$, where $x \in \mathbb{R}^2$. For $t > 0$, the perturbation is concentrated in a circle with radius $ut$ centered at $x = 0$. The advanced wavefront, given by $|x| = ut$, propagates with velocity $u$ and is followed by backward retarded wavefronts. The Huygens–Fresnel principle is thus violated, and the wave diffusion phenomenon is instead observed. To explain this behaviour, we considered the perturbation $\eta_2(t, x)$ as a convolution between the fundamental solution $\eta_3(t, x)$ and the point-like source $\delta(x_1, x_2)\mathbb{1}(x_3)\delta(t)$, as follows:

$$\eta_2(t, x_1, x_2) = \eta_3(t, x_1, x_2, x_3) * \delta(x_1, x_2)\mathbb{1}(x_3)\delta(t). \qquad (37)$$

The perturbation generated by source $\delta(x_1, x_2)\mathbb{1}(x_3)\delta(t)$ propagates as a cylindrical wave of equation $|x| \leq ut$ with an advanced wavefront perpendicular to the direction of the $x_3$-axis. After passing through a point $x$, which is external to $\Omega$, this perturbation is maintained indefinitely. In fact, at a given point $x_0 \in \mathbb{R}^2$ at instant $t > 0$, the perturbation generated by source $\delta(x_1, x_2)\mathbb{1}(x_3)\delta(t)$ will come from points that belong to the $x_3$-axis of the sphere of equation $|x - x_0|^2 + x_3{}^2 = u^2 t^2$ . It follows that for $t < |x_0|/u$, point $(0, x_0)$ will remain at rest, and for $t > |x_0|/u$, the perturbations will arrive from the points belonging to set $S = \{0, \pm(u^2 t^2 - |x - x_0|^2)^{1/2}\}$. What we have just argued is the geometric interpretation of the continuation of the fundamental solution $\eta_3(t, x)$ in space $\mathcal{D}(\mathbb{R}^2)$ that we dealt with in Section 4. This ansatz is thus useful for studying transitions from diffractive to diffusive regimes involving instantaneous sources [33–34].

Finally, let us consider the case $n = 1$. The elementary perturbation $\eta_1(t, x)$ is generated by a point-like source $\delta(x)\delta(t)$, where $x \in \mathbb{R}$. For $t > 0$, the perturbation is concentrated on the segment $-ut \leq x \leq ut$. However, two wavefronts will propagate at the same speed in straight lines in opposite directions. In this case, the advanced wavefront will be followed by backward wavefronts, generating the phenomenon of wave diffusion.

## 7. Concluding Discussion

Distribution theory rigorously justifies all the *tricks* in which chemists and physicists are engaged. Consider the Dirac delta function, the differentiation of non-differentiable functions, and the use of divergent series and integrals in quantum field theory. In the present paper, we applied Schwartz distribution theory to a classical wave equation, proposing an effective model for studying problems involving the propagation and interaction of waves generated by instantaneous sources of any form. Given their complexity, such problems are often solved by numerical methods [35–37]. However, we propose an analytical approach whereby the solution for a wave equation is obtained from the superposition of elementary perturbations generated by point-like sources that interfere with each other according to the structure of the instantaneous source. This method can be applied to study the propagation of perturbations in deforming interfaces. The latter represent a discontinuous medium in which a wave propagates and can be analytically treated only within the framework of distribution theory [38-40]. In particular, knowing the form of the waves emerging from the interfaces makes it possible to obtain information about the interfaces' structures. Each wave is the convolution of one of the fundamental solutions for the associated equation with the function that describes the interface. Therefore, it is sufficient to apply the deconvolution process to obtain the generalised function representing the interface and to obtain information on its structure. Possible applications of this approach range from photonics to chemical waves, with a potentially interesting application concerning the branch of crystallography that deals with the resolution of the structures of imperfect crystals using neutron diffractometric techniques [41]. In this case, the deconvolution of the scattered waves by the lattice would allow us to obtain function F, giving information on imperfections in the crystal. However, the most obvious analogy between the crystallographic resolution methods and the model proposed in this study is represented by the Patterson function, which is used to analyse crystals formed by molecules containing heavy atoms [42]. The Patterson function is defined as $P(u) = \int_V \rho(r)\rho(r+u)dr$, where the integral is calculated for the lattice volume. The ordinary function $\rho(r)$ represents the electron density. In function $P(u)$, all the atoms of the original structure $\rho(r)$ are moved by the same vector $u$ to obtain the corresponding $\rho(r+u)$. The products $\rho(r)\rho(r+u)$ are performed, and their contributions are added. The latter will differ from zero when the point of the translated structure coincides with the atom of the original one. This condition occurs when the $u$ vector coincides with an interatomic vector. If we compare the function $P(u)$ with solution $\omega(x) = \int f(\xi)\delta(x-\xi)d\xi$, we realise that $f(\xi)$ plays the role of the original lattice structure and $\delta(x-\xi)$ that of the translated lattice. Therefore, the convolution $f(x) * \delta(x)$ will be non-zero only when a point $\delta(x-\xi)$ coincides with a lattice vector. Compared to Patterson's function, the advantage of our method is that $f(x)$ is a general function better suited to describing a structure characterised by discontinuities. These are examples of how some theories and calculation methods can be revisited in the framework of generalised functions to describe phenomena that intrinsically deal with discontinuous or non-differentiable functions. Paraphrasing Dirac, if a theory describing a natural phenomenon is correct, it indeed has a place in the vastness of distribution theory.

## Declarations

### Competing interests
The author declare do not have any competing interests.

### Ethical approval
Not applicable